\documentclass[a4paper,11pt]{article}
\usepackage{amssymb,amsmath,bm}  
\usepackage{graphics,graphicx}           
\usepackage{array,booktabs}               
\usepackage{authblk}                          

\usepackage[linktocpage,
colorlinks = true,
linkcolor = blue,
urlcolor  = blue,
citecolor = red,
anchorcolor = blue]{hyperref}

\usepackage{epsfig}

\numberwithin{equation}{section}

\title{Skyrme-Faddeev Lagrangian from reformulated Georgi-Glashow model}
\author[1]{Ahmad Mohamadnejad\thanks{a.mohamadnejad@ut.ac.ir}}
\affil[1]{Young Researchers and Elite Club, Islamshahr Branch, Islamic Azad University, Islamshahr 3314767653, Iran}
\date{\today}

\begin{document}

\baselineskip 0.65 cm

\maketitle

\begin{abstract}
Recently we proposed a decomposition for fields of the Georgi-Glashow model and interpreted Cho's decomposition as a result of some constraints on Georgi-Glashow's fields.
Now, using the decomposition form that Faddeev and Niemi proposed, we introduce a simple and novel method to derive the Skyrme-Faddeev Lagrangian from the reformulation of the Georgi-Glashow model with an extra constraint.
As we showed before, this extra constraint leads to appearance of both vortices and monopoles.
\end{abstract}

\section{Introduction} \label{sec1}

The Skyrme-Faddeev model \cite{SF1,SF2} is a (3+1)-dimensional modified O(3) sigma-model. It includes a term quartic in derivatives to provide a preferred scale and it presents
topological solitons, called Hopfions, which have the form of stable closed strings which form knots and links. This model has links to low energy QCD. The theory possesses topological
knot solitons classified by the homotopy group
$ \Pi_{3} (CP^{1}) = Z $, i.e., by the topological Hopf charge. It is suggested to interpret such knot solutions as color electric and color magnetic glueball states \cite{Cho1,Niemi1,Cho2,Cho3}. One should notice that derivation of a explicit expression for a low energy effective action from the basics of QCD is an extremely difficult problem \cite{Cho4,Cho5,Cho51,Gies}. There exists a number of various extended Skyrme-Faddeev models where some exact and numeric knot and vortex solutions have been found \cite{Ferreira1,Ferreira2,Ferreira3,Shi}.

Despite many attempts, the final description of the low energy sector of the Yang-Mills theory
still remains a challenge for theoretical physicists. A fascinating approach to this problem
is based on some non-perturbative methods which involve soliton solutions.
The importance of
solitons in particle physics has increased meaningfully since it became clear that they could play
a role as suitable normal modes in the description of the low energy regime for some physical
theories. For example, there are theories that describe color confinement in low energy QCD via vortices and monopoles.

One can extract both vortices and monopoles by filed decomposition in Yang-Mills theories \cite{Ahmad1,Ahmad2}.
Yang-Mills field decomposition was first suggested by Cho \cite{Cho6,Cho7} and has been
developed by Faddeev and Niemi \cite{FN1,FN2,FN3,FN4} and Shabanov \cite{Shabanov1,Shabanov2}. Note that decomposition of fields does not change the physical content of the original theory. However, some aspects of the original theory are hidden, and they can be revealed by choosing the appropriate variables. Indeed, this is the philosophy of the field decomposition proposed by Cho, Faddeev, Niemi, and others. This method enables one to
describe and understand some aspects of a given theory by separating the
contributions of the topological defects in a gauge-invariant way. In the Cho method,
the additional magnetic symmetry is established and it leads to a decomposition of
the Yang-Mills field with six dynamical degrees of freedom. One can show that
magnetic monopoles emerge in this decomposition, and, based on their condensations,
quark confinement is explained \cite{Grigorio,Ahmad3} in the framework of the dual superconductor
picture proposed in 70s \cite{Nambu,Creutz,Hooft1,Mandelstam}. 
Faddeev and Niemi also proposed a decomposition for $ SU(N) $ Yang-Mills field. Their decomposition differs from Cho decomposition; the first one is on-shell decomposition, and it contains only physical degrees of freedom, while the second one is off-shell decomposition which is implemented with a pure topological field.
However, Faddeev-Niemi decomposition does not describe full QCD \cite{Evslin,Niemi2}. Based on their decomposition, Faddeev and Niemi made an interesting conjecture that the
Skyrme-Faddeev action could be interpreted as an effective action for QCD.

In this paper, after reviewing the reformulated Georgi-Glashow model \cite{Ahmad4}, we study
the extraction of effective Lagrangians involving soliton solutions such as vortices, monopoles and Hopfions.
These effective Lagrangians are derived by considering some constraints
called vacuum conditions.
We had studied the appearance of both vortices and monopoles in the framework of Faddeev-Niemi decomposition previously,
but we reiterate them here for the sake of completeness.
The unprecedented part of this paper is allocated to deriving Skyrme-Faddeev Lagrangian from
the reformulated Georgi-Glashow model. Note that, the Skyrme-Faddeev Lagrangian as well as its generalizations had been already obtained
from pure Yang-Mills theory by complicated procedures \cite{Cho4,FN1,Shabanov1}.
However, in the framework of Georgi-Glashow model with new variables and considering some constraints that leads to appearance of both vortices and monopoles, we show that one can simply obtain Skyrme-Faddeev Lagrangian without any intricacy.
Therefore, we conclude that Skyrme-Faddeev model is an effective model of reformulated Georgi-Glashow model with some constraints.
Moreover, Skyrme-Faddeev model supports knotlike soliton solutions. Therefore, solitons
such as vortices, monopoles and Hopfions can appear in the reformulated Georgi-Glashow model.

This paper is organized as follows: In Sec. 2, we review reformulation of
Georgi-Glashow model. In Sec. 3, we show how vortices and
monopoles appear in reformulated Georgi-Glashow model with some constraints.
In Sec. 4 we obtain Skyrme-Faddeev model as a constrained Georgi-Glashow model.
 Hence, Skyrme-Faddeev Lagrangian can
be viewed as an effective Lagrangian of SU(2) Georgi-Glashow model
and knotlike solitons as well as vortices and monopoles exist in this theory.
Finally our conclusion comes in Sec. 5.

\section{Reformulation of Georgi-Glashow model} \label{sec2}

In \cite{Ahmad4}, we proposed a decomposition for SU(2) Georgi-Glashow fields and obtained a Lagrangian based on new variables.
The SU(2) Georgi-Glashow model has the following classical Lagrangian,
\begin{align}
L &= \frac{1}{2} (\partial_{\mu} \bm{\phi}  + g  \textbf{A}_{\mu} \times \bm{\phi})^{2} \nonumber\\
&-\frac{1}{4} (\partial_{\mu} \textbf{A}_{\nu} - \partial_{\nu} \textbf{A}_{\mu} + g \textbf{A}_{\mu} \times \textbf{A}_{\nu})^{2} \nonumber\\
&  - \frac{\lambda}{4} (\bm{\phi} \, . \, \bm{\phi}  -  \nu^{2})^{2} ,
 \label{eq01}
\end{align}
where $ g $ and $ \lambda $ are gauge and scalar coupling constants, respectively, and the constant $ \nu $ is the scalar field vacuum expectation value.
 
The following constraints on the classical fields minimize the energy:
\begin{align}
\partial_{\mu} \bm{\phi}  + g  \textbf{A}_{\mu} \times \bm{\phi} &=  0  , \label{eq02}  \\
\bm{\phi} \, . \, \bm{\phi}  &=  \nu^{2} .  \label{eq03}
\end{align}
We refer to Eq. (\ref{eq02}) and Eq. (\ref{eq03}) as vacuum conditions.
These conditions are the same as the 't Hooft-Polyakov monopole solution constraints in the boundary \cite{Hooft2,Polyakov}.
However, we generalize these vacuum conditions for the bulk as well as the boundary.

The Higgs field $ \bm{\phi} $ has a magnitude and a direction in internal space and can be written as
\begin{equation}
\bm{\phi} =   \phi \, \textbf{m} , \label{eq04}
\end{equation}
where $ \phi $ is the magnitude of $ \bm{\phi} $, and $ \textbf{m} $ is a unit vector ($ \textbf{m} \, . \, \textbf{m} = 1 $).
From Eq. (\ref{eq04}) one gets
\begin{equation}
\triangledown_{\mu} \bm{\phi} =  (\partial_{\mu} \phi) \textbf{m} + \phi \triangledown_{\mu} \textbf{m}     , \label{eq05}
\end{equation}
where $ \triangledown_{\mu} $ is the covariant derivative operator ($ \triangledown_{\mu} = \partial_{\mu} + g \textbf{A}_{\mu} \times $). Therefore
\begin{align}
 \triangledown_{\mu} \textbf{m} &= \partial_{\mu} \textbf{m} + g \textbf{A}_{\mu} \times \textbf{m}  , \nonumber  \\
&\Rightarrow \textbf{m} \times \triangledown_{\mu} \textbf{m} = \textbf{m} \times  \partial_{\mu} \textbf{m}
+g \textbf{A}_{\mu} - g (\textbf{A}_{\mu} . \textbf{m}) \textbf{m} , \nonumber  \\
&\Rightarrow \textbf{A}_{\mu} =  (\textbf{A}_{\mu} . \textbf{m}) \textbf{m} + \frac{1}{g} \partial_{\mu} \textbf{m} \times \textbf{m} +
\frac{1}{g} \textbf{m} \times \triangledown_{\mu} \textbf{m}  . \label{eq06}
\end{align}
Introducing two new fields, $ A_{\mu} $ and $ \textbf{X}_{\mu} $ such as the following
\begin{align}
A_{\mu} &=  \textbf{A}_{\mu} . \textbf{m}  , \nonumber  \\
\textbf{X}_{\mu} &=  \frac{1}{g} \textbf{m} \times \triangledown_{\mu} \textbf{m} \, , \quad  (\textbf{X}_{\mu} \, . \, \textbf{m} = 0) , \label{eq07}
\end{align}
we have
\begin{equation}
\textbf{A}_{\mu} = A_{\mu}  \textbf{m} + \frac{1}{g} \partial_{\mu} \textbf{m} \times \textbf{m} +  \textbf{X}_{\mu}   , \label{eq08}
\end{equation}
which is Cho's extended decomposition.

We can rewrite the Higgs field $ \bm{\phi} $ and Yang-Mills field $ \textbf{A}_{\mu} $ based on new fields:
\begin{align}
\bm{\phi} &=   \phi \, \textbf{m}  , \nonumber  \\
\textbf{A}_{\mu} &= A_{\mu}  \textbf{m} + \frac{1}{g} \partial_{\mu} \textbf{m} \times \textbf{m} +  \textbf{X}_{\mu} , \label{eq09}
\end{align}
where there are some constraints:
\begin{equation}
\textbf{m} \, . \, \textbf{m}  = 1 \, , \quad \quad \textbf{m} \, . \, \textbf{X}_{\mu} =0. \label{eq10}
\end{equation}

Substituting new variables (\ref{eq09}) in the Georgi-Glashow equations:
\begin{align}
 \triangledown_{\nu} \textbf{F}^{\mu\nu} &= g \bm{\phi}  \times   \triangledown^{\mu} \bm{\phi}   , \nonumber\\
\triangledown_{\mu} \triangledown^{\mu} \bm{\phi}  &=  - \lambda \bm{\phi}  (\bm{\phi} \, . \, \bm{\phi}  -  \nu^{2}), \label{eq11}
\end{align}
we have
\begin{align}
& \triangledown_{\nu} \textbf{F}^{\mu\nu} = g^{2}  \phi^{2} \, \textbf{X}^{\mu}   , \label{eq12} \\
 & - \lambda \phi  (\phi^{2}  -  \nu^{2}) \textbf{m})= (\partial_{\mu} \partial^{\mu} \phi) \textbf{m} \nonumber \\ 
 &  + 2 g (\partial_{\mu} \phi) (\textbf{X}^{\mu} \times \textbf{m})
+g \phi \triangledown_{\mu} (\textbf{X}^{\mu} \times \textbf{m}) . \label{eq13}
\end{align}
Equation (\ref{eq13}) can be decomposed to two equations:
\begin{align}
\partial_{\mu} \partial^{\mu} \phi &= g^{2} \phi \, \textbf{X}_{\mu} \, . \, \textbf{X}^{\mu}  - \lambda \phi  (\phi^{2}  -  \nu^{2}) , \label{eq14} \\
\triangledown_{\mu} [\phi^{2} \textbf{X}^{\mu}] &= 0 .  \label{eq15}
\end{align}
Eq. (\ref{eq15}) can be obtained from Eq. (\ref{eq12}):
\begin{equation}
 \triangledown_{\nu} \textbf{F}^{\mu\nu} = g^{2}  \phi^{2} \textbf{X}^{\mu} \, \Rightarrow \,
\triangledown_{\mu} \triangledown_{\nu} \textbf{F}^{\mu\nu} = g^{2} \triangledown_{\mu}  [\phi^{2} \textbf{X}^{\mu}] = 0   . \label{eq16}
\end{equation}
Therefore, there are two independent equations:
\begin{align}
 \triangledown_{\nu} \textbf{F}^{\mu\nu} &= g^{2}  \phi^{2} \, \textbf{X}^{\mu} , \nonumber  \\
\partial_{\mu} \partial^{\mu} \phi &= g^{2} \phi \, \textbf{X}_{\mu} \, . \, \textbf{X}^{\mu}  - \lambda \phi  (\phi^{2}  -  \nu^{2}) .  \label{eq17}
\end{align}
These equations are derivable from the following Lagrangian,
\begin{align}
L &= \frac{1}{2}  (\partial_{\mu} \phi) (\partial^{\mu} \phi) + \frac{1}{2} g^{2} \phi^{2} \textbf{X}_{\mu} \, . \, \textbf{X}^{\mu} , \nonumber \\
& - \frac{1}{4} \textbf{F}_{\mu\nu} \, . \, \textbf{F}^{\mu\nu} - \frac{\lambda}{4} (\phi^{2}  -  \nu^{2})^{2}   , \label{eq18}
\end{align}
which is a reformulated Georgi-Glashow Lagrangian \cite{Ahmad4}.

The Euler-Lagrange equations of Lagrangian  (\ref{eq18}) are
\begin{align}
\textbf{m} \, . \, \triangledown_{\nu} \textbf{F}^{\mu\nu} &= 0 , \label{eq19} \\
\triangledown_{\nu} \textbf{F}^{\mu\nu} &= g^{2}  \phi^{2} \, \textbf{X}^{\mu} , \label{eq20}  \\
\partial_{\mu} \partial^{\mu} \phi &= g^{2} \phi \, \textbf{X}_{\mu} \, . \, \textbf{X}^{\mu}  - \lambda \phi  (\phi^{2}  -  \nu^{2}) ,  \label{eq21}
\end{align}
and variation with respect to $ \textbf{m} $ gets a trivial identity.
Furthermore, considering the constraint (\ref{eq10}), Eq. (\ref{eq19}) can be derived from Eq. (\ref{eq20}).

The equations of motion of the reformulated Georgi-Glashow model, Eqs. (\ref{eq20}) and  (\ref{eq21}),
are the same as the primary ones, Eq.  (\ref{eq17}). Therefore, our reformulation does not change the dynamics of the Georgi-Glashow model,
at least at the classical level.

\section{Vortices and monopoles in reformulated Georgi-Glashow model} \label{sec3}

In this section, we consider the following constraints on the field strength tensor $ \textbf{F}_{\mu\nu} $
\begin{equation}
\textbf{F}_{\mu\nu}=F_{\mu\nu} \textbf{m} , \label{eq22}
\end{equation}
where $ F_{\mu\nu} $ is a colorless tensor and $ \textbf{m} $ gives the internal direction at each space-time point.
In the following, we will show that this constraint leads to the appearance of vortices and monopoles
 \cite{Ahmad1}.
Consider a special form of $ \textbf{X}^{\mu} $ that Faddeev and Niemi proposed:
\begin{equation}
\textbf{X}_{\mu} = \frac{\rho}{g^{2}} \partial_{\mu} \textbf{m} + \frac{\sigma}{g^{2}} \textbf{m}\times\partial_{\mu} \textbf{m}, \label{eq23}
\end{equation}
where $ \rho $ and $ \sigma $ are real scalar fields. Applying these new variables, constraint (\ref{eq22}) leads to
\begin{equation}
\begin{aligned}
\partial_{\mu} \rho - g A_{\mu} \sigma &= 0, \\
\partial_{\mu} \sigma + g A_{\mu} \rho &= 0. \label{eq24}
\end{aligned}
\end{equation}
The above equations can be written in a concise form
\begin{eqnarray}
D_{\mu} \varphi  =  0, \label{eq25}
\end{eqnarray}
where $ D_{\mu} $ is the covariant derivative, $ D_{\mu} = \partial_{\mu} + i g A_{\mu} $, and $ \varphi $
is a complex scalar field, $ \varphi = \rho + i \sigma $.

The constraint on the field $ \varphi $  via Eq.  (\ref{eq25}) leads
to the appearance of vortices in the theory.
Eq. (\ref{eq25}) shows how the fields  $ \rho $, $ \sigma $, and $ A_{\mu} $ depend to each other.
A trivial solution for Eq. (\ref{eq25}) is
\begin{eqnarray}
\rho = \sigma = 0, \label{eq26}
\end{eqnarray}
which leads to the Cho's decomposition. The field strength tensor for Cho's restricted theory is
\begin{equation}
\textbf{F}_{\mu\nu}=\lbrace \partial_{\mu}  A_{\nu} - \partial_{\nu}  A_{\mu}  - \frac{1}{g} \textbf{m} .  (\partial_{\mu} \textbf{m} \times \partial_{\nu} \textbf{m}) \rbrace \textbf{m} \label{eq27}
\end{equation}

The constraint (\ref{eq25}) restricts
the Abelian U(1) gauge field and leads to the appearance of string-like (vortex) objects.
Equation (\ref{eq25}) implies
\begin{equation}
\partial_{\mu} (\rho^{2} + \sigma^{2}) = 0 \, \Rightarrow \, \varphi^{\ast} \varphi= \rho^{2} + \sigma^{2} = a^{2} , \label{eq28}
\end{equation}
where $ a $ is a constant. The non-zero value of $ a $ brings some fascinating difference between the decomposition
here and the original decomposition of Cho and plays an essential role in the appearance of vortices.
Equation (\ref{eq25}) can be solved exactly for $ A_{\mu} $
\begin{eqnarray}
A_{\mu} = \frac{1}{g a^{2}} (\sigma \partial_{\mu} \rho - \rho \partial_{\mu} \sigma). \label{eq29}
\end{eqnarray}
In Eq. (\ref{eq29}), the Abelian gauge field  $ A_{\mu} $ is decomposed to the scalar fields $\sigma$ and $\rho$.

The field strength tensor $ \textbf{F}_{\mu\nu} $ can be written in terms of electric and magnetic field strength tensors, $ G_{\mu\nu} $ and $ B_{\mu\nu} $, respectively
\begin{equation}
\textbf{F}_{\mu\nu} = (G_{\mu\nu} + B_{\mu\nu}) \textbf{m} , \label{eq30}
\end{equation}
where
\begin{align}
G_{\mu\nu} &= \partial_{\mu} A_{\nu} - \partial_{\nu} A_{\mu}, \\
B_{\mu\nu} &= -\frac{1}{g^{\prime}} \textbf{m} .  (\partial_{\mu} \textbf{m} \times \partial_{\nu} \textbf{m}) = (1-\frac{\rho^{2} + \sigma^{2} }{g^{2}}) H_{\mu\nu} , \label{eq32} \\
H_{\mu\nu} &= -\frac{1}{g} \textbf{m} .  (\partial_{\mu} \textbf{m} \times \partial_{\nu} \textbf{m}). \label{eq31-33}
\end{align}
Since $ \rho^{2} + \sigma^{2} = a^{2} $, the contributions of $ \rho $ and $ \sigma $ are included in the new coupling $ g^{\prime} $ , where
\begin{equation}
\frac{1}{g^{\prime}} =  \frac{1}{g} - \frac{a^{2}}{g^{3}}. \label{eq34}
\end{equation}
We take $ a \leqslant g $ in order to have $ g^{\prime} \geqslant 0 $.

Note that both vortices and monopoles can appear in the Faddeev-Niemi decomposition with the constraint (\ref{eq25}).
There are two different and independent fields, $ \phi $ and $ \textbf{m} $, in the reformulated theory which can have different boundary conditions and generate different topological structures, i.e., vortices and monopoles. For the field $ \phi = \rho + i \sigma $, the boundary is $ S^1 $ and it is responsible for the appearance of vortices, while for the field $ \textbf{m} $, the boundary is  $ S^2 $ and it is responsible for the appearance of the monopoles.
Vortices appear in this theory as topological objects.
To study vortices, we take the boundary of the space to be a circle at infinity,
denoted by $ S^{1}_{R} $.
Vortices are characterized by the homotopy class of a mapping $ \Pi_{1} (S^{1}) $
of the spatial circle $ S^{1}_{R} $ to the coset space $ S = U(1) $ of the internal space.
To define this mapping, one needs a two components scalar field in the theory,
at least on $ S^{1}_{R} $. The scalar fields $ \rho $ and $ \sigma $ in decomposition (\ref{eq29}) can be used to define the
mapping $ \Pi_{1} (S^{1}) $. We define the topological charge by
the homotopy class of the mapping $ \Pi_{1} (S^{1}) $ given by $ (\rho,\sigma) $
\begin{equation}
(\rho,\sigma); \quad S^{1}_{R} \rightarrow S^{1} = U(1). \label{eq35}
\end{equation}
The homotopy class $ \Pi_{1} (S^{1}) $ defined by
the following ansatz describes the vortex with a unit flux tube
\begin{equation}
(\rho,\sigma)=a\frac{\overrightarrow{r}}{r}=a (cos(\varphi),sin(\varphi)) , \label{eq36}
\end{equation}
where here $ \varphi $ is the azimuthal circular coordinate of $ S^{1}_{R} $ and $ r $ is the distance from $ z $ axis in the cylindrical coordinate system.
Using Eq. (\ref{eq36}) in Eq. (\ref{eq29}) one obtains
\begin{eqnarray}
A_{\mu} = - \frac{1}{g} \partial_{\mu} \varphi,  \nonumber\\ &&
\hspace{-30mm} \Longrightarrow A_{r} = A_{z} = 0 \, , \, \,\,\,\, A_{\varphi} = - \frac{1}{gr} . \label{eq37}
\end{eqnarray}
The magnetic field $ \overrightarrow{B} $ can be obtained as the following
\begin{eqnarray}
\overrightarrow{B} = \overrightarrow{\nabla} \times \overrightarrow{A} =
\widehat{r} (\frac{1}{r} \frac{\partial A_{z}}{\partial \varphi} -  \frac{\partial A_{\varphi}}{\partial z}) +
\widehat{\varphi} ( \frac{\partial A_{r}}{\partial z} -  \frac{\partial A_{z}}{\partial r} )  \nonumber\\ &&
\hspace{-62mm} + \widehat{k} \frac{1}{r} ( \frac{\partial (r  A_{\varphi})}{\partial r} -  \frac{\partial A_{r}}{\partial \varphi} ) = 0  . \label{eq38}
\end{eqnarray}
In general we have
\begin{align}
(\rho,\sigma)&=a (cos(\alpha),sin(\alpha)) \nonumber \\
& \Rightarrow A_{\mu} = - \frac{1}{g} \partial_{\mu} \alpha
\Rightarrow G_{\mu\nu} = 0 . \label{eq39}
\end{align}
The above calculations are true at every place in space, but not on the z axis where $ r=0 $.
The magnetic flux passing through the closed curve is not zero:
\begin{eqnarray}
\phi_{B} = \int_{S} \overrightarrow{B} .
 \overrightarrow{ds} = \int_{S} (\overrightarrow{\nabla} \times \overrightarrow{A}) .
 \overrightarrow{ds} = \oint_{A} \overrightarrow{A} . \overrightarrow{dl}  \nonumber\\ &&
\hspace{-64mm} = \int^{2\pi}_{0}  - \frac{1}{gr} rd\varphi = - \frac{2\pi}{g}. \label{eq40}
\end{eqnarray}

It shows that on the z axis the magnetic field is singular as well as $ A_{\varphi} $ in Eq. (\ref{eq37}).
Therefore, although the magnetic field is zero everywhere, there exists an infinite magnetic field on the z axis, responsible for the magnetic flux of Eq. (\ref{eq40})
which is an evidence of a vortex lying on the z axis. One can obtain the magnetic field:
\begin{eqnarray}
\int_{S} \overrightarrow{B} . \overrightarrow{ds} = \int^{R}_{0} \int^{2\pi}_{0} B \, rd\theta \, dr =  - \frac{2\pi}{g} \nonumber\\ &&
\hspace{-55mm} \Longrightarrow \int^{R}_{0} B \, rdr  = -\frac{1}{g} \nonumber\\ &&
\hspace{-55mm} \Longrightarrow B = -2 \frac{\delta(r)}{gr}, \label{eq41}
\end{eqnarray}
Notice that the string
tension of the vortex is infinite.
To get vortices with finite string tension, one should consider that eq. (\ref{eq25}) is valid just for the boundary of the vortex solution, $ r \rightarrow \infty $, not the bulk.

We can find out all the homotopically inequivalent classes of the mapping (\ref{eq35}) and the corresponding
vortex configurations by the following replacement
\begin{eqnarray}
\varphi \rightarrow n \, \varphi. \label{eq42}
\end{eqnarray}
Then we have
\begin{equation}
A_{r} = A_{z} = 0 \, , \, \,\,\,\, A_{\varphi} = - \frac{n}{gr} \quad  (r \rightarrow \infty) , \label{eq43}
\end{equation}
and magnetic flux is:
\begin{eqnarray}
\phi_{B} = - \frac{2\pi n}{g}. \label{eq44}
\end{eqnarray}

In addition to the vortices, monopoles can also emerge in SU(2) Yang-Mills theory. According to Eqs. (\ref{eq30}) and (\ref{eq32}) we have:
\begin{eqnarray}
\textbf{F}_{\mu\nu} = (G_{\mu\nu} + B_{\mu\nu}) \textbf{m} ,
\end{eqnarray} \label{eq45}
where
\begin{eqnarray}
B_{\mu\nu} = \mu (\varphi ^{\ast} \varphi) H_{\mu\nu} , \label{eq46}
\end{eqnarray}
and
\begin{eqnarray}
\mu (\varphi ^{\ast} \varphi) = (1-\frac{\varphi ^{\ast} \varphi}{g^{2}}). \label{eq47}
\end{eqnarray}
$ \mu (\varphi ^{\ast} \varphi) $ is a parameter characteristic of the medium; we call it ''vacuum permeability".
We have
\begin{eqnarray}
0 \leqslant \mu (\varphi ^{\ast} \varphi) \leqslant 1 . \label{eq48}
\end{eqnarray}

The topological magnetic charges can be described by the homotopy class
of the mapping $ \Pi_{2}(S^{2}) $ of the two-dimensional sphere $ S^{2}_{R} $ to the coset space
$ S^{2} = SU(2) / U(1)$ of the internal space.
To obtain the magnetic field from $ H_{\mu\nu} $, we choose a hedgehog configuration for $ \textbf{m} $
\begin{equation}
\textbf{m}=\frac{r^{a}}{r}=
\begin{pmatrix}
\sin{\alpha} \, cos{\beta} \\
\sin{\alpha} \, sin{\beta} \\
\thickspace cos{\alpha}
\end{pmatrix}
\end{equation} \label{eq49}
where
$$ \alpha = \theta, \quad
\beta = m \varphi. $$
$ \theta $ and $ \varphi $ are the angular spherical coordinates of $ S^{2}_{R} $, and $ m $
is an integer number. The magnetic intensity $ H $ can be obtained,
\begin{eqnarray}
\overrightarrow{H} = H \widehat{r}, \nonumber\\ &&
\hspace{-18mm} H = H_{\theta\varphi} = - \frac{1}{g} \textbf{m} . (\partial_{\theta} \textbf{m} \times \partial_{\varphi} \textbf{m})
 =  - \frac{m}{g} \frac{1}{r^{2}} . \label{eq50}
\end{eqnarray}

The ''vacuum permeability" depends on the value of the condensed field.
We have
\begin{equation}
\overrightarrow{B} = \mu (\varphi ^{\ast} \varphi) \overrightarrow{H} . \label{eq51}
\end{equation}
The vacuum behaves like a superconducting medium in which the scalar field  $ \varphi $ is condensate.
The magnetic field $ \overrightarrow{B} $ depends on the ''vacuum permeability" and in the Higgs phase it goes to zero.
thus the vacuum which is structured by the Higgs field $ \varphi $ does not allow the presence of the magnetic field except in vortex form
and these vortices can confine monopoles.

\section{The Skyrme-Faddeev model as a constrained Georgi-Glashow model} \label{sec4}

Now we focus on vacuum condition (\ref{eq03}) of the Georgi-Glashow model as the condensate phase and we derive a Lagrangian that is a generalization
of the Cho-Faddeev-Niemi Lagrangian.
The condensate phase (\ref{eq03}) in which the Higgs field takes the vacuum expectation value, $ \phi = \nu $, leads to the following effective Lagrangian:
\begin{align}
L &=  \frac{1}{2} g^{2} \nu^{2} \, \textbf{X}_{\mu} \, . \, \textbf{X}^{\mu} - \frac{1}{4} \textbf{F}_{\mu\nu} \, . \, \textbf{F}^{\mu\nu} \nonumber \\
 &=  \frac{1}{2} g^{2} \nu^{2} \, \textbf{X}_{\mu} \, . \, \textbf{X}^{\mu} - \frac{1}{4} \widehat{\textbf{F}}_{\mu\nu} \, . \, \widehat{\textbf{F}}^{\mu\nu} \nonumber \\
 &   - \frac{1}{4} ( \widehat{\triangledown}_{\mu} \textbf{X}_{\nu} -  \widehat{\triangledown}_{\nu} \textbf{X}_{\mu}).( \widehat{\triangledown}^{\mu} \textbf{X}^{\nu} -  \widehat{\triangledown}^{\nu} \textbf{X}^{\mu}) \nonumber \\
 &  - \frac{g^{2}}{4} (\textbf{X}_{\mu} \times \textbf{X}_{\nu}).(\textbf{X}^{\mu} \times \textbf{X}^{\nu})
 - \frac{g}{2} \widehat{\textbf{F}}_{\mu\nu} \, . \, (\textbf{X}^{\mu} \times \textbf{X}^{\nu})     . \label{eq52}
\end{align}
where
\begin{align}
\widehat{\textbf{F}}_{\mu\nu} &= \partial_{\mu} \widehat{\textbf{A}}_{\nu} -  \partial_{\nu} \widehat{\textbf{A}}_{\mu} +
g \widehat{\textbf{A}}_{\mu} \times \widehat{\textbf{A}}_{\nu} = (G_{\mu\nu} + H_{\mu\nu})  \textbf{m} , \nonumber \\
\widehat{\textbf{A}}_{\mu} &= A_{\mu} \textbf{m} + \frac{1}{g} \partial_{\mu} \textbf{m} \times \textbf{m}, \nonumber \\
\widehat{\triangledown}_{\mu} \textbf{X}_{\nu} &= \partial_{\mu} \textbf{X}_{\nu} + g \widehat{\textbf{A}}_{\mu} \times \textbf{X}_{\nu} . \label{eq53}
\end{align}

In this phase, $ \textbf{X}_{\mu} $ gets mass
\begin{equation}
m_{\textbf{X}} = g \, \nu . \label{eq54}
\end{equation}
Considering Eq.  (\ref{eq15}), we have
\begin{equation}
\triangledown_{\mu}  \textbf{X}^{\mu} = 0 \quad \Rightarrow \quad \widehat{\triangledown}_{\mu}  \textbf{X}^{\mu} = 0 .  \label{eq55}
\end{equation}
This condition was compelled on the Cho extended decomposition in order to compensate for the two extra degrees introduced by $ \textbf{m} $  \cite{Cho4}.

By Substituting Eq. (\ref{eq23}) in Lagrangian (\ref{eq52}), we generalize the Faddeev-Niemi Lagrangian
\begin{align}
L &= \frac{1}{2} \frac{\nu^{2}}{g^{2}} \, \varphi ^{\ast} \varphi \, \,   \partial_{\mu} \textbf{m} \, . \,  \partial^{\mu} \textbf{m} -\frac{1}{4} G_{\mu\nu} G^{\mu\nu}  \nonumber\\
&  + \frac{1}{2g^{4}} ( \partial_{\mu} \textbf{m}.\partial_{\nu} \textbf{m} - \eta_{\mu\nu} \partial_{\lambda} \textbf{m}.\partial^{\lambda} \textbf{m} )
(D^{\mu}\varphi)^{\ast}  (D^{\nu}\varphi) \nonumber\\
& + \frac{i}{2g^{3}} H_{\mu\nu} (D^{\mu}\varphi)^{\ast}  (D^{\nu}\varphi)  - \frac{1}{2} H_{\mu\nu} G^{\mu\nu} (1-\frac{\varphi ^{\ast} \varphi}{g^{2}})  \nonumber\\
& - \frac{1}{4}  H_{\mu\nu} H^{\mu\nu} (1-\frac{\varphi ^{\ast} \varphi}{g^{2}})^{2} ,  \label{eq56}
\end{align}
where the first term is added to the Faddeev-Niemi Lagrangian and it leads to new results.
Now, we show how the Skyrme-Faddeev Lagrangian can be derived from the above Lagrangian by considering constraint  (\ref{eq25})
which is responsible for the appearance of both vortices and monopoles.
This constraint yields eq. (\ref{eq28}) and eq. (\ref{eq39}):
\begin{equation}D_{\mu} \varphi = 0 \Rightarrow \left\{
  \begin{array}{lr}
    \varphi^{*} \varphi = a^{2} \\
    G_{\mu \nu} = 0
  \end{array}
\right. \label{eq57}
\end{equation}
Therefore, Lagrangian (\ref{eq56}) reduce to:
\begin{eqnarray}
L = \frac{1}{2} \frac{\nu^{2}}{g^{2}} \, a^{2} \, \,   \partial_{\mu} \textbf{m} \, . \,  \partial^{\mu} \textbf{m}
 - \frac{1}{4}  H_{\mu\nu} H^{\mu\nu} (1-\frac{a^{2}}{g^{2}})^{2} .  \label{eq58}
\end{eqnarray}
The Lagrangian (\ref{eq58}) is the same as Skyrme-Faddeev Lagrangian:
\begin{eqnarray}
L = M^{2} (\partial_{\mu} \textbf{m})^{2}
 - \frac{1}{e^{2}} (\partial_{\mu} \textbf{m} \times \partial_{\nu} \textbf{m})^{2}  \label{eq59}
\end{eqnarray}
where
\begin{align}
M^{2} &= \frac{1}{2} \frac{\nu^{2}}{g^{2}} \, a^{2} , \\
\frac{1}{e^{2}} &= \frac{1}{4 g^{2}}(1-\frac{a^{2}}{g^{2}})^{2} .  \label{eq60-61}
\end{align}

Therefore, the Skyrme-Faddeev Lagrangian which describes knotlike solitons can be interpreted as an effective Lagrangian of the condensate phase of
our reformulation of the Georgi-Glashow model.

Note that (\ref{eq59}) is the unique local and Lorentz-invariant Lagrangian for the unit vector $ \textbf{m} $ which is at most quadratic in time derivatives so that it admits a Hamiltonian interpretation and includes all such terms that are either relevant or peripheral in the vacuum limit. It is also noticeable that in four dimensions the Lagrangian (\ref{eq59}) fails to be perturbatively renormalizable in the ultraviolet limit. However, since it is anticipated to describe the physical excitations of a SU(2) Yang-Mills-Higgs theory in the vacuum strong coupling limit,
lack of perturbative renormalizability should not present a
problem provided that we can interpret (\ref{eq59}) adequately. Indeed, it has been already established that in $ 3+1 $ dimensions the classical Lagrangian (\ref{eq59}) describes stable knotlike solitons (for example see \cite{SF2,Battye}). This suggests that a proper route to its quantization should be based on the investigation of the quantum mechanical properties of these solitons.

\section{Conclusion} \label{sec5}

We have recently proposed a reformulation of the Georgi-Glashow model that is
equivalent to the Georgi-Glashow model at least at the
classical level. We define some constraints for this reformulated model called vacuum conditions
and derive effective Lagrangians that describe some hidden aspects of the theory.
The results presented in this paper reveal that there are many limits associated with different constraints
on reformulated Georgi-Glashow model that contains knotlike solitons as well as vortices and monopoles.
One of this constraints leads to appearance of both vortices and monopoles.
Applying this constraint together with the condensate phase constraint of the reformulated Georgi-Glashow model,
leads us to derive Skyrme-Faddeev Lagrangian that contains knotlike solitons.

\section*{Acknowledgements}
This work is supported financially by the
Young Researchers and Elite Club of Islamshahr Branch of
Islamic Azad University.

\end{document}